\documentclass[numeric]{aa}
\usepackage[dvipsnames]{xcolor}

\usepackage{multirow}
\usepackage{graphicx}
\usepackage{txfonts}
\usepackage{hyperref}
\usepackage{xcolor}
\usepackage{subcaption}
\usepackage{bm}
\usepackage{fontawesome5}
\hypersetup{colorlinks,citecolor=blue}
\hypersetup{colorlinks=true,linkcolor=magenta,filecolor=magenta,    urlcolor=blue}

\begin{document} 

\newcommand{\affubb}{Department of Physics, Babeș-Bolyai University, Kogălniceanu Street, Cluj-Napoca, 400084, Romania}
\newcommand{\affkhazar}{Research Center of Astrophysics and Cosmology, Khazar University, 41 Mehseti Street, AZ1096, Baku, Azerbaijan}
\newcommand{\affunina}{Dipartimento di Fisica ``E. Pancini'', Università degli Studi di Napoli ``Federico II'', Complesso Universitario di Monte Sant’ Angelo, Edificio G, Via Cinthia, I-80126, Napoli, Italy}
\newcommand{\affinfn}{Istituto Nazionale di Fisica Nucleare (INFN), Sezione di Napoli, Via Cinthia 9, I-80126, Napoli, Italy}
\newcommand{\affssm}{Scuola Superiore Meridionale, Via Mezzocannone 4, I-80134, Napoli, Italy}
\newcommand{\affstx}{Department of Physics, St. Xavier's College, Tribhuvan University, Kathmandu, Nepal}
\newcommand{\affzjnu}{Department of Physics, Zhejiang Normal University, Jinhua 321004, People's Republic of China}
\newcommand{\affhit}{School of Physics, Harbin Institute of Technology, Harbin 150001, People's Republic of China}

\title{Thermodynamic Parameterizations of Dark Energy in Light of DESI DR2 BAO}

\author{
Himanshu Chaudhary\inst{1,2} \and
Salvatore Capozziello\inst{3,4,5} \and
Anil Kandel\inst{6} \and
G.~Mustafa\inst{7} \and
Chengxun Yuan\inst{8}
}

\institute{
$^{1}$\,\affubb\\
$^{2}$\,\affkhazar\\
$^{3}$\,\affunina\\
$^{4}$\,\affinfn\\
$^{5}$\,\affssm\\
$^{6}$\,\affstx\\
$^{7}$\,\affzjnu\\
$^{8}$\,\affhit
}

\date{}

\abstract{{The physical origin of cosmic acceleration remains one of the central unresolved problems of modern cosmology. Although the $\Lambda$CDM model successfully describes a wide range of cosmological observations, its interpretation in terms of a strictly constant vacuum energy is affected by long-standing theoretical challenges, motivating the search for physically grounded alternatives. In this work, we develop a thermodynamic parametrization of dark energy (TPDE)  by combining the second law of thermodynamics with the kinematic evolution of the Universe. By imposing physically motivated constraints on the cosmic deceleration parameter and reconstructing its evolution through a Padé approximant, we derive a simple one-parameter extension of the $\Lambda$CDM scenario in which the late-time accelerated expansion emerges naturally from thermodynamic principles rather than from an assumed cosmological constant. We constrain the model using the Planck 2018 legacy CMB temperature and polarization observations, Planck 2018 CMB lensing, DESI DR2 BAO measurements, and the Pantheon$+$, DES-Dovekie, and Union3 Type Ia supernova compilations. The analysis shows that the present-day thermodynamic parameter remains tightly constrained around the $\Lambda$CDM limit, implying that any departure from a cosmological constant is small with current observations. While the model does not resolve the $H_0$ tension because it preserves the sound horizon at the drag epoch, it remains in good agreement with current weak-lensing constraints on $S_8$, showing tensions below $0.2\sigma$ with KiDS-Legacy (Shear) and below $1.5\sigma$ with DES~Y6. Despite providing maximum-likelihood fits, comparable to those of $\Lambda$CDM, Bayesian model comparison consistently yields a weak-to-moderate preference for the TPDE scenario. These results demonstrate that thermodynamic reconstruction provides a simple, physically motivated, and observationally competitive alternative to the standard $\Lambda$CDM paradigm, offering a promising framework for investigating the fundamental origin of dark energy by forthcoming high-precision cosmological observations.}}

\maketitle

The physical mechanism driving the late-time accelerated expansion of the Universe remains one of the central unresolved problems of modern cosmology. Since its discovery through the observed dimming of distant Type Ia supernovae \citep{SupernovaSearchTeam:1998fmf,SupernovaCosmologyProject:1998vns}, identifying the origin of this acceleration has become a major focus of theoretical and observational research. The simplest and most successful dark energy candidate is the cosmological constant $\Lambda$, corresponding to the specific case $w=-1$. A universe with the cosmological constant $\Lambda$ as dark energy is commonly referred to as the $\Lambda$CDM cosmology model, where CDM denotes cold dark matter. The standard $\Lambda$CDM cosmological model provides the most successful framework for describing the today observed large-scale  of the Universe. With a minimal set of parameters, it yields a consistent account of observations spanning a wide range of epochs and physical scales, including anisotropies in the cosmic microwave background (CMB) \citep{hinshaw2013nine, Planck:2018vyg, ACT:2023kun}, the large-scale distribution of matter, BAO \citep{DESI:2024mwx}, and luminosity-distance measurements from Type Ia supernovae (SNIa) \citep{SupernovaSearchTeam:1998fmf,SupernovaCosmologyProject:1998vns}. Despite its empirical success, the $\Lambda$CDM model is confronted with persistent theoretical difficulties, particularly the cosmological constant problem \citep{Weinberg:1988cp} and the coincidence problem \citep{zlatev1999quintessence}.

Beyond these conceptual difficulties, the increasing precision of cosmological observations has revealed persistent discrepancies between independent datasets, thereby placing additional pressure on the internal consistency of the $\Lambda$CDM framework. The most prominent of these is the Hubble tension \citep{di2021realm,cai2026hubble}, which refers to the statistically significant difference between the present-day expansion rate, inferred indirectly from early-Universe observations, and that determined from direct late-Universe measurements. In particular, the value of $H_0$ obtained from Planck CMB data \citep{aghanim2020planck} under the assumption of $\Lambda$CDM differs from the local distance-ladder determination reported by the SH0ES collaboration \citep{riess2022comprehensive}, with the discrepancy having been reported at a significance exceeding $5\sigma$. A joint analysis of CMB observations from Planck, SPT-3G, and ACT \citep{aghanim2020planck,SPT-3G:2025bzu,ACT:2020gnv,AtacamaCosmologyTelescope:2025blo}, compared with the homogeneous set of direct late-time measurements compiled by the $H_0$ Distance Network (H0DN), yields a discrepancy of approximately $7.1\sigma$ \citep{H0DN:2025lyy}. This result constitutes one of the most statistically significant manifestations of the Hubble tension reported thus far. See also \citep{  Capozziello:2023ewq,Capozziello:2024stm} for a discussion.

A separate but closely related tension concerns the amplitude of matter fluctuations at late times, conventionally characterized by $S_8 \equiv \sigma_8\sqrt{\Omega_m/0.3}$. Weak-lensing and other large-scale-structure measurements generally favor lower values of $S_8$ than those inferred from Planck Cosmic Microwave Background (CMB) observations within the standard $\Lambda$CDM framework~\citep{Pantos:2026koc,DES:2026zjp,wright2025kids}. This discrepancy may point to a broader inconsistency between cosmological constraints derived from early- and late-Universe observations.

Beyond these persistent tensions, recent BAO measurements from the Dark Energy Spectroscopic Instrument (DESI) have provided important new clues regarding the possible dynamical nature of dark energy. The first-year DESI BAO measurements, when combined with CMB data and the Pantheon$^{+}$, Union3, and DESY5 Type~Ia supernova samples, exhibited a preference for evolving dark energy over a cosmological constant, with statistical significances ranging from $2.5\sigma$ to $3.9\sigma$ depending on the supernova compilation employed~\citep{DESI:2024mwx}. This trend became more pronounced with the second DESI data release (DESI DR2) in 2025. In combination with CMB and the Pantheon$^{+}$, Union3, and DESY5 datasets, the updated BAO measurements yielded preferences over the $\Lambda$CDM model at the $3.1\sigma$, $2.8\sigma$, $3.8\sigma$, and $4.2\sigma$ levels, respectively~\citep{abdul2025desi}. Nevertheless, subsequent reanalyses of the DESY5 supernova sample, particularly those based on the revised DES-Dovekie dataset, reduced the corresponding significance to approximately $3.2\sigma$, compared with the previously reported $4.2\sigma$ preference obtained using the original DES-SN5YR analysis~\citep{popovic2025dark}. Despite the sensitivity of the statistical preference to the adopted supernova treatment, the collective findings suggest that a strictly constant vacuum-energy component may provide an incomplete description of late-time cosmic acceleration.

Motivated by these observational indications, we investigate whether thermodynamic principles can provide model-independent constraints on the cosmic expansion history without specifying the microscopic origin of dark energy. In particular, we reconstruct the early and late-time evolution of a spatially flat Friedmann--Lemaître--Robertson--Walker (FLRW) Universe by translating the second law of thermodynamics and the requirement of asymptotic equilibrium into direct restrictions on cosmological kinematics. In the exact de~Sitter limit, characterized by a constant Hubble
parameter, the apparent and cosmological horizons
are degenerate \citep{Wang:2005pk}. Away from this limit, for a dynamical cosmological evolution characterized by
$w\neq -1$, the apparent horizon generally does not coincide with the cosmological event horizon. The apparent horizon is defined quasi-locally by the spacetime geometry and remains well defined in a dynamical FLRW Universe. It consequently provides a natural setting for imposing thermodynamic constraints on the cosmic expansion history. 

The second law of thermodynamics requires the entropy associated with the horizon to evolve monotonically with respect to a suitable cosmological evolution variable, such that $dS\geq0$. Furthermore, the attainment of thermodynamic equilibrium requires the entropy to become asymptotically
concave, $d^{2}S<0$, as $z\rightarrow-1$.
 These two conditions restrict the admissible behavior
of the deceleration parameter,
$q=-\ddot{a}/(aH^{2})=-1-\dot{H}/H^{2}$, and thereby constrain the
expansion history at early, present, and future epochs. 
To obtain a stable and model-independent reconstruction of the cosmic
expansion history, we introduce a flexible parametrization of the
deceleration parameter based on rational Pad\'e approximants. Specifically,
we adopt the $(0,1)$ Pad\'e form, whose numerator is a zeroth-order
polynomial and whose denominator is linear in the chosen cosmological
variable. This lowest-order rational approximation remains well behaved at
high redshift and avoids the convergence limitations commonly associated
with truncated Taylor expansions. The choice is guided by the requirement that dark energy became dynamically negligible at early times, so that the standard matter- and radiation-dominated epochs are recovered. In the present construction, this behavior is naturally realized through $q_{\rm de}\rightarrow 0$ as $z\rightarrow\infty$.
Since the parametrization is formulated
directly in terms of the deceleration parameter, no particular cosmological
model is assumed a priori. The resulting expansion history is then used to
reconstruct the effective properties of dark energy and to identify possible departures from the standard $\Lambda$CDM scenario.

The present analysis builds upon the thermodynamic parametrization introduced in Ref.~\citep{Capozziello:2022jbw}, where the background expansion was constrained using model-independent observational Hubble data and the Pantheon Type~Ia supernova compilation, while the evolution of matter perturbations was examined using the Gold-2017 compilation of
weak-lensing and redshift-space-distortion measurements. Here, we subject the thermodynamically constrained parametrization to a substantially broader observational test by combining the full CMB likelihood with DESI DR2 BAO measurements and three independent Type~Ia supernova compilations: Pantheon$^{+}$, Union3, and DES-Dovekie.
These dataset combinations provide complementary constraints over a broad redshift range and allow us to assess the robustness of the reconstructed expansion history with respect to the adopted supernova sample.

The paper is organized as follows. Section~\ref{sec_2} presents the thermodynamic framework and the adopted parametrization of the cosmic expansion history. Section~\ref{sec_3} describes the observational datasets,
methodology, and parameter estimation procedure. Section~\ref{sec_4} presents the resulting cosmological constraints and discusses their implications for
the expansion history and the inferred properties of dark energy. Finally, Section~\ref{sec_5} summarizes the main conclusions of this work.

\section{Theoretical Background}\label{sec_2}
Thermodynamic principles have become increasingly important in modern cosmology because of their broad applicability \citep{Pavon:2012qn}. In particular,  thermodynamically motivated dark-energy scenarios have been widely explored as viable alternatives for explaining the observed cosmic acceleration\citep{Radicella:2011qpl,Silva:2011pc}.
Complementarily, the expansion history can be characterized kinematically
through the deceleration parameter without specifying a particular
dark-energy model. Here, we combine these two approaches by relating
thermodynamic constraints on the cosmological entropy to the kinematic
evolution of the deceleration parameter, thereby reconstructing the past,
present, and future expansion history without assuming a specific
dark-energy model a priori.

\subsection{Constraints from cosmological entropy}
Thermodynamic evolution imposes two fundamental conditions on an isolated macroscopic system. The first follows from the second law of thermodynamics, according to which the entropy must be non-decreasing as the system evolves, implying $dS \geq 0$.
 Second, as the system approaches
thermodynamic equilibrium, the entropy must tend toward a maximum.
Consequently, its growth progressively slows and the entropy becomes a
concave function near the equilibrium configuration. Applied to cosmology,
this requirement implies that in the asymptotic future, corresponding to
$z \rightarrow -1$, the entropy should satisfy $d^{2}S < 0$.

In a FLRW spacetime, the entropy budget at late times is dominated by the
entropy associated with the apparent horizon, which provides a natural
thermodynamic boundary for the cosmological system. The radius of the apparent
horizon is given by
\begin{equation}
\tilde{r}_{h}
=
\left(
H^{2}+\frac{k}{a^{2}}
\right)^{-1/2},
\label{eq:apparent_horizon}
\end{equation}
where $H=\dot{a}/a$ is the Hubble parameter, $a$ is the scale factor, and
$k$ denotes the spatial curvature. For the spatially flat case considered in
this work, $k=0$, and the apparent-horizon radius therefore reduces to
$\tilde{r}_{h}=H^{-1}$.

Neglecting quantum corrections, the entropy associated with the apparent
horizon obeys the Bekenstein--Hawking area law,
\begin{equation}
S_{h}
=
\frac{A_{h}}{4G}
=
\frac{\pi}{G H^{2}},
\label{eq:horizon_entropy}
\end{equation}
where $A_{h}=4\pi\tilde{r}_{h}^{\,2}$. Differentiating $S_h$ with respect to
cosmic time yields
\begin{equation}
\dot{S}_{h}
=
-\frac{2\pi}{G H^{3}}\dot{H},
\label{eq:entropy_derivative}
\end{equation}

Using the definition of the deceleration parameter,
\begin{equation}
q
=
-1-\frac{\dot{H}}{H^{2}},
\label{eq:deceleration_parameter}
\end{equation}
or equivalently $\dot{H}=-H^{2}(1+q)$, the entropy-production rate can be
expressed as
\begin{equation}
\dot{S}_{h}
=
\frac{2\pi}{G H}(1+q),
\label{eq:entropy_production}
\end{equation}

Hence, for an expanding Universe with $H>0$, fulfillment of the second law,
$\dot{S}_{h}\geq 0$, implies the condition $q\geq -1$.

To characterize the approach toward thermodynamic equilibrium, we require the apparent-horizon entropy to asymptotically approach a finite and positive maximum value. In the far-future limit, this implies
$\dot{S}_{h}\rightarrow 0$, together with the concavity condition
$\ddot{S}_{h}<0$. From the Bekenstein-Hawking relation
$S_{h}=\pi/(G H^{2})$, the asymptotic condition
$S_{h}\rightarrow S_{h,\infty}$, with
$0<S_{h,\infty}<\infty$, leads to
$H\rightarrow H_{\infty}
=
\sqrt{\frac{\pi}{G S_{h,\infty}}}>0,
\label{eq:asymptotic_H}$

Using Eq.~(\ref{eq:entropy_production}), since the prefactor $2\pi/(G H_{\infty})$ remains finite and nonzero in the asymptotic regime, the condition $\dot{S}_{h}\rightarrow 0$ can only be satisfied if $1+q\rightarrow 0$. Consequently, the thermodynamic equilibrium condition selects the asymptotic behavior
$q\rightarrow -1,
\label{eq:q_asymptotic}
$

Thus, the de~Sitter limit is not imposed a priori, but instead follows from the requirement that the horizon entropy approach a finite maximum in the far future.

A further differentiation of the entropy-production rate with respect to cosmic time yields
\begin{equation}
\ddot{S}_{h}
=
\frac{2\pi}{G}
\left[
\frac{\dot{q}}{H}
+
(1+q)^{2}
\right],
\label{eq:entropy_second_derivative}
\end{equation}

Once the asymptotic condition $q\rightarrow -1$ is reached, the second term
vanishes, and the entropy curvature reduces to
\begin{equation}
\ddot{S}_{h}
\sim
\frac{2\pi}{G H}\dot{q},
\label{eq:entropy_second_asymptotic}
\end{equation}

Since $H\rightarrow H_{\infty}>0$, the requirement of asymptotic
thermodynamic equilibrium, $\ddot{S}_{h}<0$, therefore implies
$\dot{q}<0,
\label{eq:qdot_condition}
$
Hence, the combined requirements of the second law and asymptotic
thermodynamic equilibrium imply that the deceleration parameter remains bounded by $q\geq -1$ and approaches the de~Sitter value $q=-1$ from above in the far-future limit.

\subsection{Thermodynamically constrained reconstruction of the expansion history}

We consider a spatially flat Universe composed of pressureless matter and an effective dark energy component, whose dynamics are governed by the Friedmann equations. Throughout this work, we use natural units with $8\pi G = c = 1$.
\begin{align}
&H^2=\dfrac{1}{3}\left(\rho_m+\rho_{de}\right), \\
&3H^2+2\dot{H}=- p_{de}\,.
\end{align}

Rather than parametrizing the dark energy equation of state, we reconstruct the cosmic expansion through the deceleration parameter. For this purpose, we use Pad\'e approximants \citep{pade1892representation}, whose rational form provides a stable description of the expansion history over a wide redshift range and avoids the divergence problems of ordinary Taylor series \citep{visser2004jerk,gruber2014cosmographic, Aviles:2014rma, Petreca:2023nhy, Benetti:2019gmo, Capozziello:2018jya}. The general $(n,m)$ Pad\'e approximant is defined as
\begin{equation}
P_{n,m}(z)=\dfrac{\displaystyle \sum_{i=0}^n a_i z^i}{1+\displaystyle\sum_{k=1}^m b_k z^k}\,,
\end{equation}
where $a_i$ and $b_k$ are constant coefficients, while $n$ and $m$ denote the polynomial orders of the numerator and denominator, respectively \citep{baker1996pade}.

The dark energy deceleration parameter is introduced through
\begin{equation}
q_{de}=-1+\dfrac{(1+z)}{H_{de}}\dfrac{dH_{de}}{dz}\,,
\label{eq:q_de}
\end{equation}
where
\[
H_{de}^2=\frac{1}{3}\rho_{de}.
\]

To reconstruct the dark energy evolution, we adopt the lowest-order $(0,1)$ Pad\'e approximation,
\begin{equation}
q_{de}(z)=\dfrac{q_{de,0}}{1+q_1 z}\,,
\label{eq:q_de2}
\end{equation}
where $q_{{\rm de},0}$ represents the current value of the dark-energy deceleration parameter at $(z=0)$. This parametrization naturally satisfies $q_{de}\rightarrow0$ as $z\rightarrow\infty$, ensuring that dark energy becomes negligible at early times while introducing the minimum number of additional parameters.

Integrating Eq.~(\ref{eq:q_de}) gives
\begin{equation}
H_{de}=\mathcal C(1+z)^{1+\frac{q_{de,0}}{1-q_1}}(1+q_1z)^{-\frac{q_{de,0}}{1-q_1}}\,,
\end{equation}
where $q_1\neq1$, and the normalization constant is
\begin{equation}
\mathcal C=H_0\sqrt{1-\Omega_{m0}}\,.
\end{equation}

Using $\rho_{de}=3H_{de}^2$ together with $\rho_m=3H_0^2\Omega_{m0}(1+z)^3$, the total Hubble expansion rate becomes
\begin{align}
&H^2=H_0^2\bigg[\Omega_{m0}(1+z)^3+(1-\Omega_{m0})(1+z)^{2+\frac{2q_{de,0}}{1-q_1}} \nonumber\\
&\hspace{4.5cm}\times(1+q_1z)^{-\frac{2q_{de,0}}{1-q_1}}\bigg].
\label{eq:H2}
\end{align}

The corresponding total deceleration parameter is
\begin{equation}
q=\dfrac{2q_{de,0}(1-\Omega_{m0})(1+q_1z)^{-1+\frac{2q_{de,0}}{q_1-1}}+\Omega_{m0}(1+z)^{1+\frac{2q_{de,0}}{q_1-1}}}{2\left[(1-\Omega_{m0})(1+q_1z)^{\frac{2q_{de,0}}{q_1-1}}+\Omega_{m0}(1+z)^{1+\frac{2q_{de,0}}{q_1-1}}\right]}.
\end{equation}
Imposing the asymptotic thermodynamic-equilibrium condition
$q_{\rm de} \rightarrow -1$ as $z \rightarrow -1$ fixes the remaining
Pad\'e coefficient as
\begin{equation}
q_1 = 1 + q_{{\rm de},0}.
\end{equation}
leading to
\begin{equation}
q(z)=\dfrac{2q_{de,0}(1-\Omega_{m0})(1+z+q_{de,0}z)+\Omega_{m0}(1+z)^3}{2\left[(1-\Omega_{m0})(1+z+q_{de,0}z)^2+\Omega_{m0}(1+z)^3\right]}\,.
\end{equation}

Finally, the reconstructed Hubble expansion rate takes the form
\begin{equation}
H(z)=H_0\sqrt{\Omega_{m0}(1+z)^3+(1-\Omega_{m0})(1+z+q_{de,0}z)^2}\,.
\label{eq:H}
\end{equation}

This expression represents a one-parameter extension of the $\Lambda$CDM model and reduces exactly to the standard cosmology in the limit $q_{de,0}\rightarrow-1$.

The reconstructed Hubble expansion rate also enables us to derive the effective properties of the dark energy component. For a general dynamical dark energy scenario, the normalized Hubble parameter can be written as
\begin{equation}
E(z)^2=\Omega_{m0}(1+z)^3+(1-\Omega_{m0})X(z)\,,
\end{equation}
where
\begin{equation}
X(z)=\exp{\left\{3\int_0^z \dfrac{1+w_{de}(z')}{1+z'}\, dz' \right\}}\,.
\end{equation}
Differentiating the latter relation yields
\begin{equation}
w_{\rm de}(z)
=
-1
+
\frac{1+z}{3}
\frac{d\ln X(z)}{dz}.
\end{equation}

For the reconstructed expansion history, the dark-energy contribution is given by
\begin{equation}
X(z)
=
\left[
1+\left(1+q_{{\rm de},0}\right)z
\right]^2,
\end{equation}
from which the effective dark-energy equation of state follows as
\begin{equation}
w_{\rm de}(z)
=
\frac{1}{3}
\left[
\frac{2q_{{\rm de},0}}
{1+\left(1+q_{{\rm de},0}\right)z}
-1
\right].
\end{equation}
The above expression
shows that the dark energy evolution is completely specified by the parameter $q_{de,0}$. In the limiting case $q_{de,0}=-1$, one recovers $w_{de}=-1$, corresponding to the cosmological constant.

Another quantity of interest is the effective adiabatic sound speed ($c_a^2$), which characterizes the relation between the background pressure and energy-density evolution of the total cosmic fluid and is defined as
\citep{mukhanov1992theory}
\begin{equation}
c_a^2\doteq
\dfrac{\dot{p}_{\rm tot}}{\dot{\rho}_{\rm tot}}
=-1-\dfrac{\ddot{H}}{3H\dot{H}}\,,
\end{equation}
where the second equality follows from the Friedmann equations.
Substituting Eq.~(\ref{eq:H}) into the above definition yields
\begin{equation}
c_a^2=\dfrac{1+z(1+q_{de,0})-q_{de,0}}
{-3\left[1+z(1+q_{de,0})+
\frac{3\Omega_{m0}(1+z)^2}
{2(1-\Omega_{m0})(1+q_{de,0})}\right]} \,.
\end{equation}

Although the above expression contains a factor
$(1+q_{de,0})^{-1}$, the effective adiabatic sound speed
remains finite in the limit $q_{de,0}\rightarrow-1$.
For $0<\Omega_{m0}<1$ and finite $z>-1$, we obtain
$\lim_{q_{de,0}\rightarrow-1}c_a^2(z)=0$,
which corresponds to the $\Lambda$CDM limit.

\section{Dataset and Methodology}\label{sec_3}
To infer the cosmological parameters of the TPDE model, we perform a Bayesian parameter estimation using the Markov Chain Monte Carlo (MCMC) sampler \citep{lewis2002cosmological,lewis2013efficient,neal2005taking, Capozziello:2011tj}. The sampling is carried out with the publicly available cosmological inference code {\tt Cobaya}~\citep{torrado2021cobaya}~\href{https://github.com/CobayaSampler/cobaya}{\faGithub}, while the theoretical predictions are computed using the Boltzmann solver Code for Anisotropies in the Microwave Background ({\tt CAMB})~\citep{lewis2000efficient,howlett2012cmb}~\href{https://github.com/cmbant/CAMB.git}{\faGithub}. The convergence of the markov chains is monitored using the Gelman--Rubin statistic, and the analysis is considered converged only when all chains satisfy the criterion $R-1<0.01$ \citep{gelman1992inference}. The resulting chains are analyzed with the {\tt GetDist} package~\citep{lewis2025getdist}~\href{https://github.com/cmbant/getdist}{\faGithub} to obtain the corresponding posterior distributions of the model parameters.

The cosmological parameter space of the baseline $\Lambda$CDM model is given by
$\bm{\theta}_{\Lambda\mathrm{CDM}}
=
\left\{
\Omega_{\rm cdm},
\Omega_{\rm b},
100\,\theta_{\rm MC},
\ln(10^{10}A_{\rm s}),
n_{\rm s},
\tau
\right\}.$ The TPDE model extends this parameter space by introducing the present-day thermodynamic dark energy parameter $q_{\rm de,0}$, such that $\bm{\theta}_{\rm TPDE}
=
\left\{
\bm{\theta}_{\Lambda\mathrm{CDM}},
q_{\rm de,0}
\right\}.$

To constrain the TPDE model, we consider several combinations of early- and late-time cosmological observations. In particular, we employ the Cosmic Microwave Background measurements from the \textit{Planck} 2018 legacy release, including the temperature and polarization anisotropy data together with the CMB lensing likelihood, the DESI DR2 BAO measurements, and Type Ia supernova observations from the Pantheon$+$, DES-Dovekie, and Union3 compilations. A brief description of these datasets is provided below.

\begin{itemize}
\item \textbf{Cosmic Microwave Background:} We use the \textit{Planck} 2018 legacy CMB observations, including the high-$\ell$ \texttt{Plik} TT, TE, and EE likelihoods, the low-$\ell$ \texttt{Commander} temperature likelihood, the low-$\ell$ \texttt{SimAll} polarization likelihood, and the \textit{Planck} 2018 CMB lensing likelihood~\citep{Planck:2019nip,Planck:2018nkj,Planck:2018lbu}.

\item \textbf{Type Ia Supernova :} Next, we consider three different compilations of Type Ia supernovae (SNe Ia). The first is the Pantheon$+$ sample~\citep{scolnic2022pantheon}~\href{https://github.com/PantheonPlusSH0ES/DataRelease.git}{\faGithub}, which contains 1,701 light curves from 1,550 SNe Ia covering the redshift range $0.001 \leq z \leq 2.26$. Following the standard Pantheon$+$ analysis \citep{brout2022pantheon}, we exclude supernovae with $z<0.01$ to reduce the impact of peculiar velocity uncertainties. We also consider the DES-Dovekie compilation~\citep{popovic2025dark}~\href{https://github.com/des-science/DES-SN5YR.git}{\faGithub}, consisting of 1,820 photometrically calibrated SNe Ia over the redshift range $0.02 \leq z \leq 1.14$. Finally, we include the Union3 compilation~\citep{rubin2025union}~\href{https://github.com/rubind/union3_release}{\faGithub}, which contains 2,087 SNe Ia spanning $0.05 \leq z \leq 2.26$. These three supernova datasets provide complementary constraints on the late-time expansion history of the Universe.

\item \textbf{Baryon Acoustic Oscillation :} Finally, we consider the BAO measurements from the Dark Energy Spectroscopic Instrument (DESI) Data Release~2 (DR2)~\citep{abdul2025desi}~\href{https://github.com/CobayaSampler/bao_data.git}{\faGithub}. The DESI DR2 sample combines measurements from the Bright Galaxy Sample (BGS), Luminous Red Galaxies (LRG), Emission Line Galaxies (ELG), Quasars (QSO), and the Lyman-$\alpha$ forest, covering the redshift range $0.3 \leq z \leq 2.33$. In our analysis, we use the compressed BAO likelihood based on the distance measurements $D_M/r_d$, $D_H/r_d$, and $D_V/r_d$. Here, $D_H(z)=c/H(z)$ is the Hubble distance, $D_M(z)=c\int_0^z dz'/H(z')$ is the transverse comoving distance, and $D_V(z)=\left[zD_M^2(z)D_H(z)\right]^{1/3}$ is the volume-averaged distance, while $r_d$ denotes the sound horizon at the baryon drag epoch.
\end{itemize}

We also perform a Bayesian comparison between the standard $\Lambda$CDM model and the TPDE scenario. The Bayesian evidence, $\mathcal{Z}$, is computed using the {\tt MCEvidence} package~\citep{heavens2017marginal,heavens2017no}~\href{https://github.com/yabebalFantaye/MCEvidence.git}{\faGithub} through its {\tt Cobaya} interface available in the {\tt wgcosmo} repository~\href{https://github.com/williamgiare/wgcosmo.git}{\faGithub}. To quantify the relative preference between the two models, we evaluate the logarithmic Bayes factor, $\ln B=\ln \mathcal{Z}_{\Lambda{\rm CDM}}-\ln \mathcal{Z}_{\rm TPDE}.$
The significance of the Bayesian evidence is assessed using the revised Jeffreys scale \citep{kass1995bayes,trotta2008bayes}. Specifically, $\ln B<1$ indicates inconclusive evidence, $1\leq\ln B<2.5$ corresponds to weak evidence, $2.5\leq\ln B<5$ to moderate evidence, $5\leq\ln B<10$ to strong evidence, and $\ln B\geq10$ to decisive evidence in favor of the $\Lambda$CDM model. In contrast, negative values of $\ln B$ indicate that the TPDE model is preferred over $\Lambda$CDM. The priors adopted for both the $\Lambda$CDM and TPDE models are listed in Table~\ref{tab_1}.

\begin{figure*}
\begin{subfigure}{.49\textwidth}
\includegraphics[width=\linewidth]{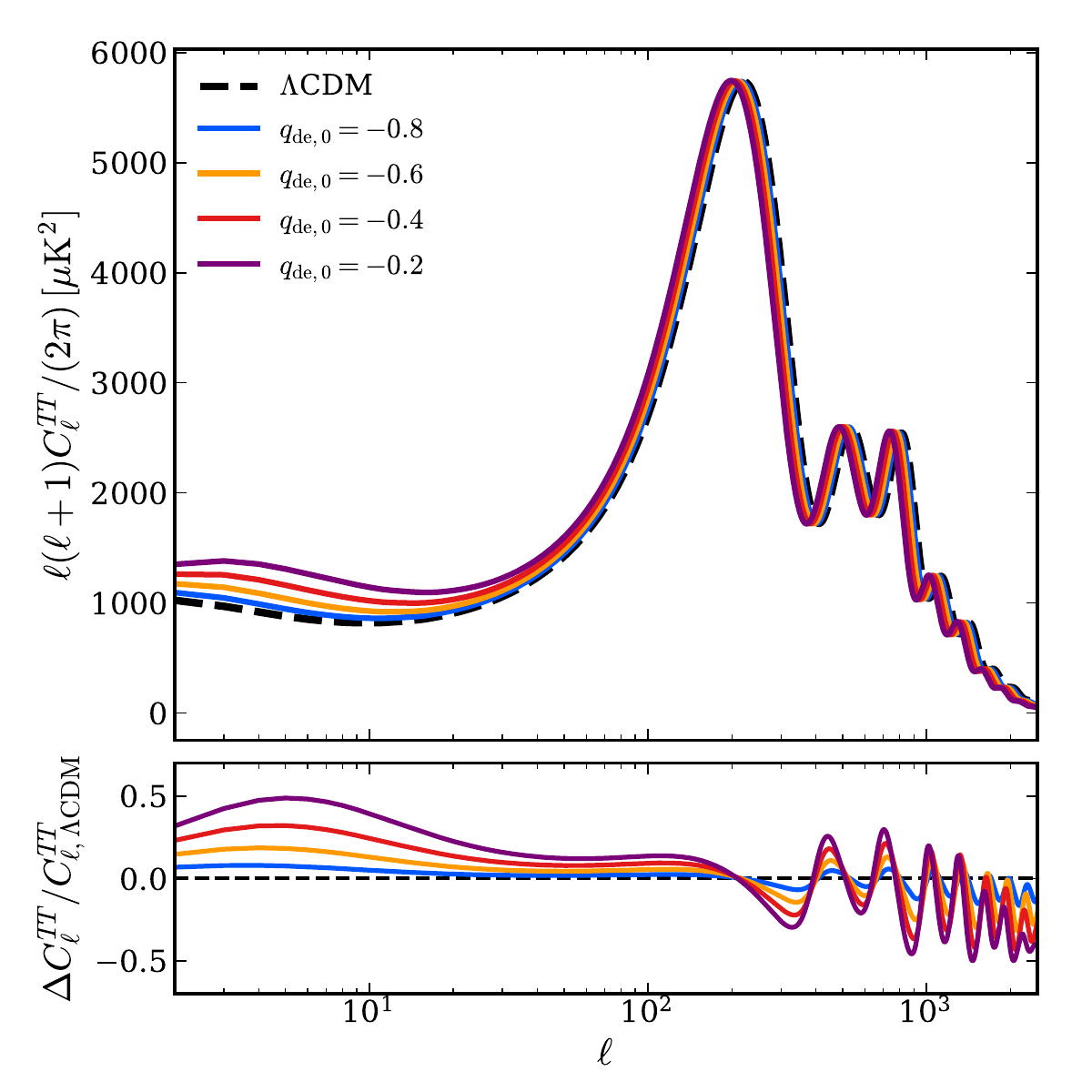}
\end{subfigure}
\hfil
\begin{subfigure}{.49\textwidth}
\includegraphics[width=\linewidth]{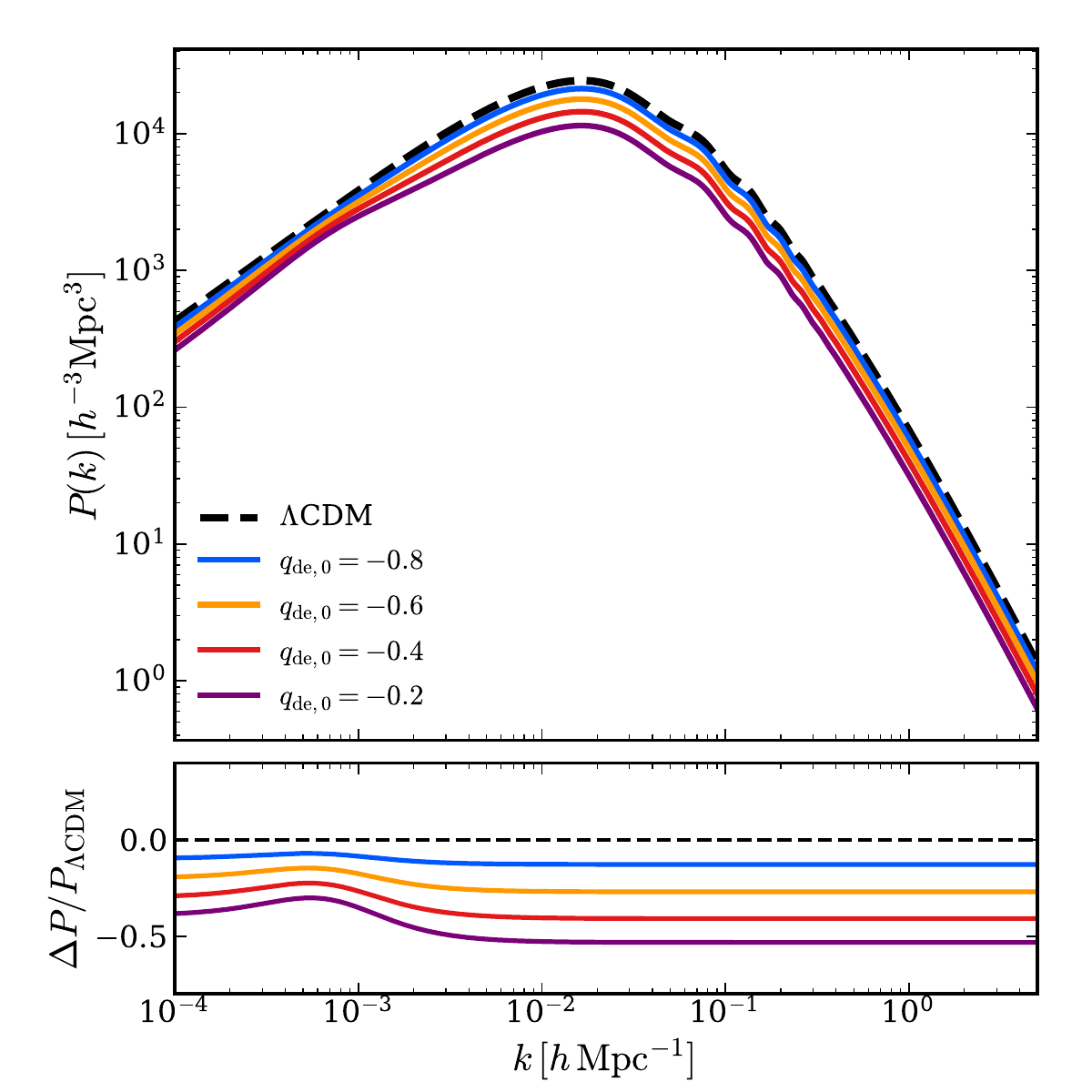}
\end{subfigure}
\caption{This figure shows the CMB temperature power spectrum, $C_\ell^{TT}$ (left panel), together with the corresponding fractional difference, $\Delta C_\ell^{TT}/C_{\ell,\Lambda{\rm CDM}}^{TT}$ (lower-left panel). The right panel shows the linear matter power spectrum, $P(k)$, and its corresponding fractional difference, $\Delta P/P_{\Lambda{\rm CDM}}$ (lower-right panel). The black dashed curve represents the $\Lambda$CDM model, while the colored curves show the predictions of the TPDE model for different values of the parameter $q_{de,0}$, with all other cosmological parameters fixed to their fiducial values.}\label{fig_1}
\end{figure*}

\begin{figure}
\centering
\includegraphics[scale=0.43]{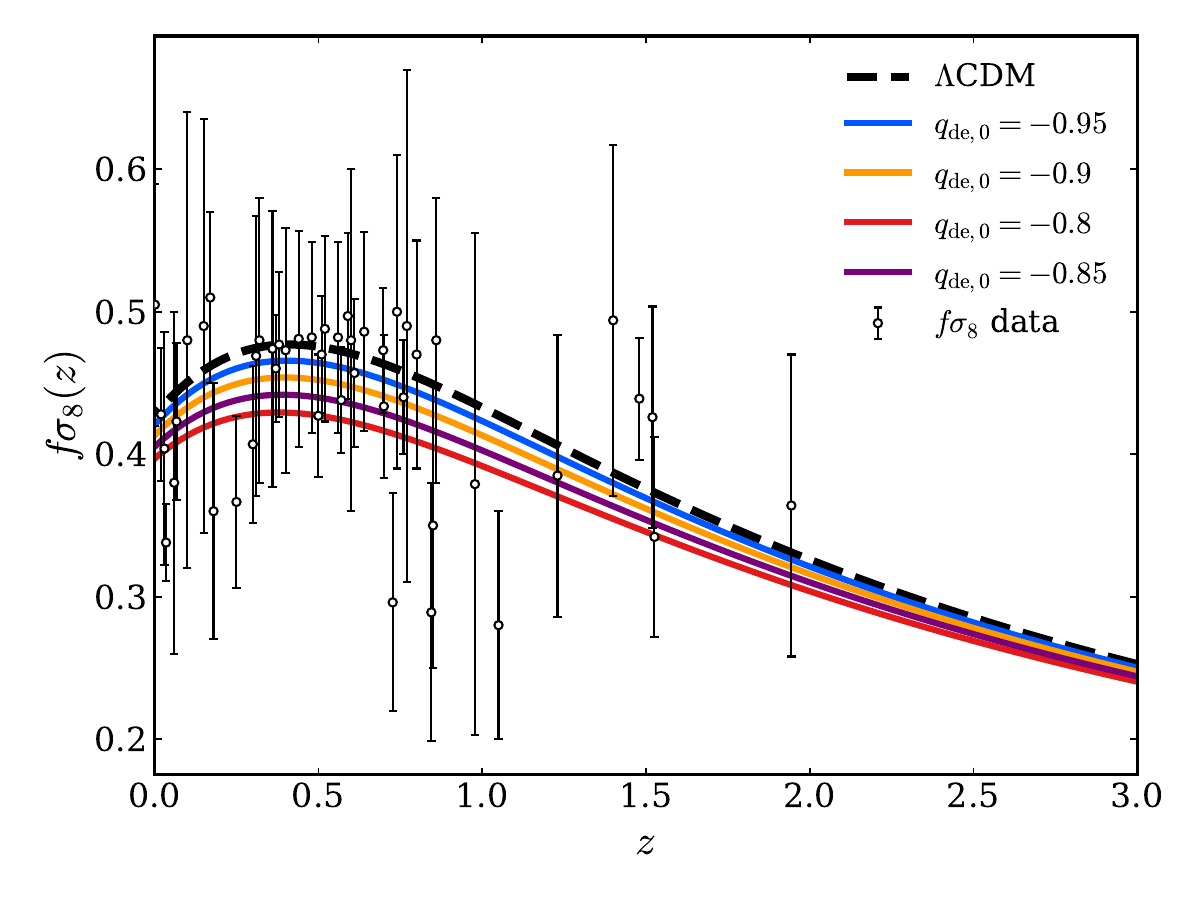}
\caption{This figure shows the evolution of the growth-rate observable $f\sigma_8(z)$ as a function of redshift for different values of the TPDE model parameter $q_{\rm de,0}$, together with the observational measurements. The black dashed curve corresponds to the $\Lambda$CDM model, while the coloured curves show the predictions of the TPDE model for different values of $q_{\rm de,0}$, with all other cosmological parameters fixed to their fiducial values.}\label{fig_2}
\end{figure}
Before constraining the TPDE model with observational data, we first explore its impact on three important cosmological observables: CMB temperature power spectrum (TT), the linear matter power spectrum ($P(k)$), and the growth-rate observable ($f\sigma_8$). Together, these observables probe both the expansion history of the Universe and the formation of cosmic structures, providing a useful way to illustrate the cosmological signatures of the TPDE model. Fig.~\ref{fig_1} presents the TT power spectrum (left panels) and the linear matter power spectrum $P(k)$ (right panels) for several representative values of the TPDE model parameter $q_{\rm de,0}$. The upper panels shows the predicted spectra, while the lower panels show their fractional differences relative to the $\Lambda$CDM model. Fig.~\ref{fig_2} presents the corresponding evolution of the growth-rate observable $f\sigma_8(z)$ together with the current observational measurements. The error bars are taken from Table~1 of Ref.~\citep{benisty2021quantifying}.

The left panels show that the TPDE model introduces only small modifications to the TT power spectrum. The largest differences occur on large angular scales ($\ell\lesssim30$), where the temperature anisotropies become progressively enhanced as $q_{\rm de,0}$ increases. The acoustic peak positions are only slightly shifted towards lower multipoles, indicating that the TPDE model has a very small effect on the acoustic scale at recombination. The amplitudes of the acoustic peaks are also only mildly modified, with the first acoustic peak becoming slightly higher for larger values of $q_{\rm de,0}$. These features are more clearly seen in the lower-left panel, where the fractional difference remains small over most multipoles but exhibits oscillatory features at high $\ell$. The oscillations follow the acoustic peak structure and mainly reflect small changes in the peak amplitudes together with the slight shift in their positions.

The right panels present the corresponding linear matter power spectrum. In contrast to the TT spectrum, the TPDE model suppresses the matter power spectrum over the entire range of scales considered. The suppression becomes stronger as $q_{\rm de,0}$ increases, while the overall shape of the spectrum remains nearly unchanged. This indicates that the TPDE model mainly reduces the overall growth of matter perturbations without introducing a significant scale dependence. The lower-right panel further shows that the fractional difference is approximately constant over most of the $k$ range, with only a small feature on the largest scales. Overall, the dominant effect of the TPDE parameter is to decrease the amplitude of the linear matter power spectrum while preserving its overall shape.

The evolution of the growth-rate observable $f\sigma_8(z)$ is also affected by the TPDE model. Compared to the $\Lambda$CDM prediction, all TPDE models predict a lower growth rate over the entire redshift range considered. The suppression becomes more pronounced for larger values of $q_{\rm de,0}$, while models with more negative values remain closer to the $\Lambda$CDM prediction. The largest differences occur at intermediate redshifts, after which the curves gradually approach the $\Lambda$CDM prediction towards higher redshifts. This behavior indicates that the TPDE parameter primarily influences the late-time growth of matter perturbations while leaving the early-time evolution almost unchanged. Within the current observational uncertainties, all TPDE models remain consistent with the available $f\sigma_8$ measurements, showing that the model is compatible with existing observations of the growth of cosmic structures.

{
\renewcommand{\arraystretch}{1}
\begin{table}[t] 
    \centering
    \begin{tabular}{|lll|}
    \hline
    parametrization & parameter & prior\\  
    \hline 
    $\mathbf{\Lambda}$\textbf{CDM} & $\Omega_\mathrm{cdm}h^2$ & $\mathcal{U}[0.001, 0.99]$ \\   
    & $\Omega_\mathrm{b}h^{2}$ & $\mathcal{U}[0.005, 0.1]$ \\
    & $100\theta_\mathrm{MC}$ & $\mathcal{U}[0.5, 10]$ \\
    & $\ln(10^{10} A_\mathrm{s})$ & $\mathcal{U}[1.61, 3.91]$ \\
    & $n_\mathrm{s}$ & $\mathcal{U}[0.8, 1.2]$ \\
    & $\tau$ & $\mathcal{U}[0.01, 0.8]$ \\
    \hline 
    \textbf{TPDE} & $q_{de,0}$ & $\mathcal{U}[-2, 0]$ \\
    \hline
    \end{tabular}
    \caption{ Parameters and priors used in the analysis. Here $\mathcal{U}[{\rm min, max}]$ denotes a uniform prior over the specified range.}\label{tab_1}
\end{table}
}


\begin{figure*}
\centering
\includegraphics[scale=0.45]{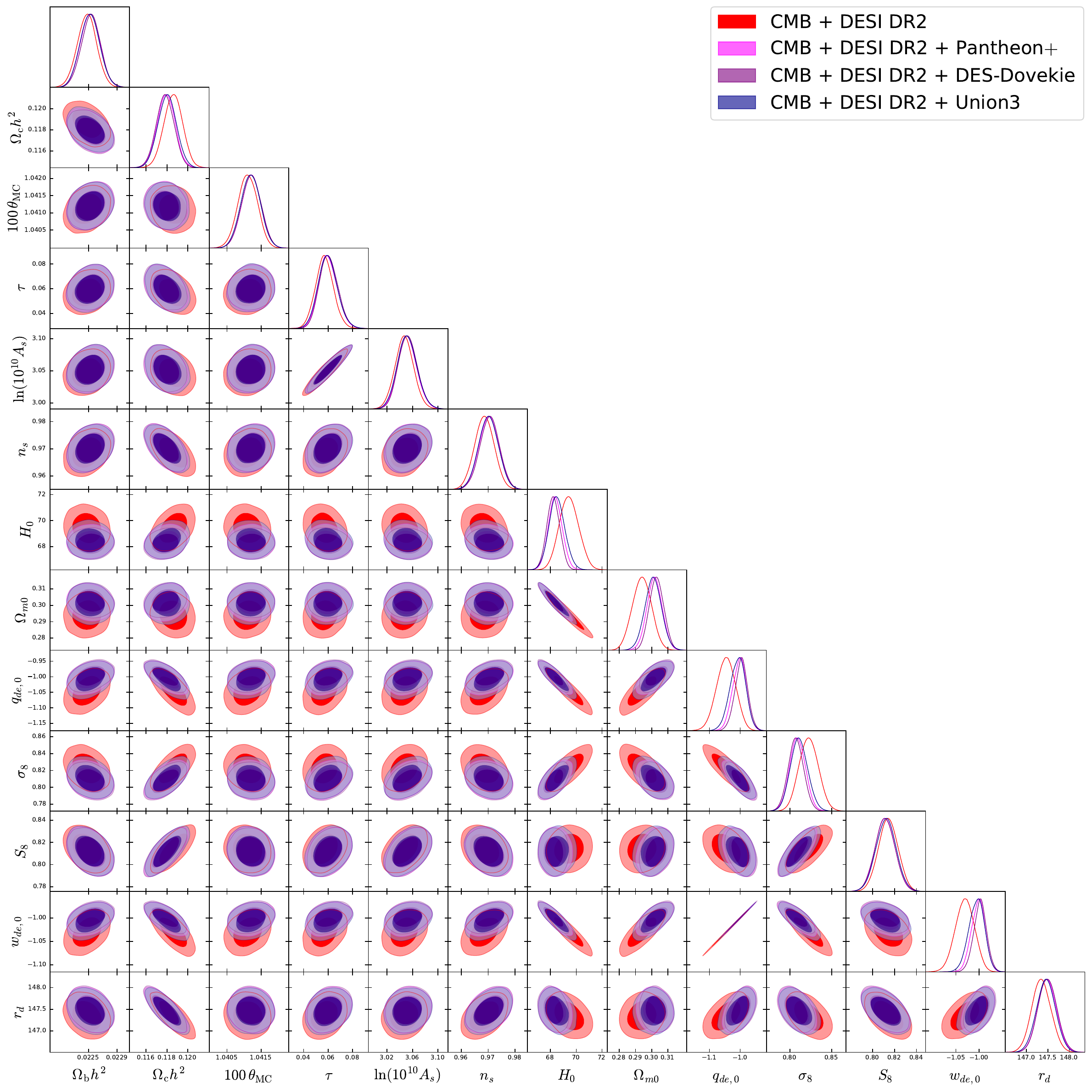}
\caption{The figure shows the corner plot of the TPDE model at the 68\% ($1\sigma$) and 95\% ($2\sigma$) confidence levels using DESI DR2 combined with the Planck 2018 CMB temperature, polarization, and lensing observations, together with the SNe~Ia datasets (Pantheon$+$, DES-Dovekie, and Union3), shown as superimposed contours for the different dataset combinations.}\label{fig_3}
\end{figure*}

\begin{figure*}
\centering
\includegraphics[scale=0.38]{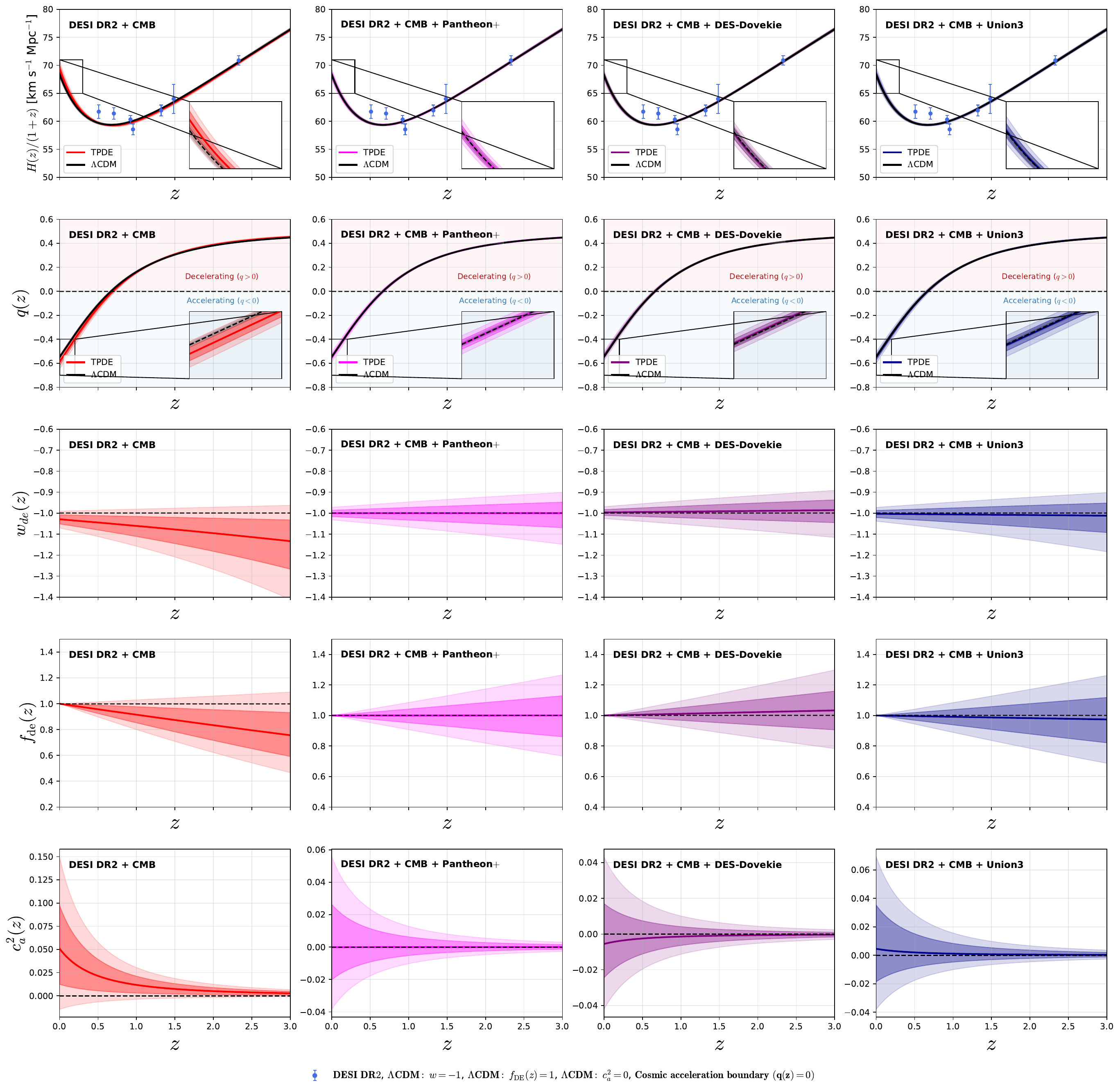}
\caption{The figure shows the evolution of $H(z)/(1+z)$ (first row), $q(z)$ (second row), $w_{\rm DE}(z)$ (third row), $f_{\rm DE}(z)$ (fourth row), and $c_a^{2}(z)$ (fifth row) for the TPDE and $\Lambda$CDM models as functions of redshift $z$. The solid lines represent the mean predictions, while the dark and light shaded regions correspond to the $1\sigma$ and $2\sigma$ confidence intervals, respectively.}\label{fig_4}
\end{figure*}

\begin{table*}[htbp]
\centering
\resizebox{\textwidth}{!}{%
\begin{tabular}{l@{\hspace{10pt}}c@{\hspace{10pt}}c@{\hspace{10pt}}c@{\hspace{10pt}}c}
\hline
\textbf{Parameters} 
& \textbf{CMB + DESI DR2} 
& \textbf{CMB + DESI DR2 + Pantheon$+$} 
& \textbf{CMB + DESI DR2 + DES-Dovekie} 
& \textbf{CMB + DESI DR2 + Union3} \\
\hline

$\Omega_{\mathrm{\text{b}}}h^2$ 
& $0.02248 \pm 0.00013$ 
& $0.02252 \pm 0.00014$
& $0.02254 \pm 0.00013$
& $0.02252 \pm 0.00014$ \\

$\Omega_{\mathrm{\text{cdm}}}h^2$ 
& $0.11860 \pm 0.00089$ 
& $0.11792 \pm 0.00085$ 
& $0.11784 \pm 0.00085$ 
& $0.11803 \pm 0.00088$  \\

$100\,\theta_{\mathrm{MC}}$ 
& $1.04111 \pm 0.00029$ 
& $1.04121 \pm 0.00028$ 
& $1.04121 \pm 0.00029$ 
& $1.04119 \pm 0.00029$  \\

$\tau_{\mathrm{reio}}$ 
& $0.0571 \pm 0.0074$ 
& $0.0599_{-0.0080}^{+0.0070}$
& $0.0605_{-0.0079}^{+0.0070}$
& $0.0597_{-0.0081}^{+0.0073}$\\

$\ln\bigl(10^{10}\,A_{\mathrm{s}}\bigr)$ 
& $3.047 \pm 0.015$ 
& $3.052 \pm 0.015$
& $3.053_{-0.016}^{+0.014}$
& $3.052 \pm 0.015$ \\

$n_{\mathrm{s}}$ 
& $0.9686 \pm 0.0036$ 
& $0.9703 \pm 0.0037$
& $0.9706 \pm 0.0036$
& $0.9701 \pm 0.0036$ \\

$q_{de,0}$ 
& $-1.046 \pm 0.031$ 
& $-1.003 \pm 0.022$
& $-0.996 \pm 0.02$ 
& $-1.008_{-0.024}^{+0.027}$ \\

\hline
$H_0 (\mathrm{km\,s^{-1}\,Mpc^{-1}})$ 
& $69.44 \pm 0.74$ 
& $68.38 \pm 0.54$
& $68.22 \pm 0.50$ 
& $68.50 \pm 0.61$ \\

$\Omega_{\mathrm{m}}$ 
& $0.2940 \pm 0.0058$ 
& $0.3018 \pm 0.0047$ 
& $0.3030 \pm 0.0044$ 
& $0.3009 \pm 0.0052$  \\

$\sigma_{8}$ 
& $0.823 \pm 0.011$ 
& $0.8091 \pm 0.0095$ 
& $0.8072 \pm 0.0092$ 
& $0.811 \pm 0.010$ \\

$S_{8}$ 
& $0.8142 \pm 0.0089$
& $0.8115 \pm 0.0091$
& $0.8112 \pm 0.0090$
& $0.8122 \pm 0.0092$ \\

$r_d\text{(Mpc)}$ 
& $147.35 \pm 0.22$ 
& $147.48 \pm 0.22$
& $147.49 \pm 0.22$
& $147.46 \pm 0.22$ \\

$w_{de,0}$ 
& $-1.031 \pm 0.020$ 
& $-1.002 \pm 0.015$
& $-0.998 \pm 0.014$ 
& $-1.005_{-0.016}^{+0.018}$ \\

\hline

$\Delta{\chi^{2}_{\text{MAP}}}$
& 0.39
& 0.57
& 2.02 
& 0.96
\\

$\ln B_{\Lambda \rm CDM,\; TPDE}$
& $-2.88$
& $-1.36$
& $-1.04$
& $-1.49$
\\

\hline
\end{tabular}
}
\caption{This table shows the constraints on the TPDE parameters, presenting the mean values and their uncertainties at the 68\% ($1\sigma$) confidence level, obtained using DESI DR2 combined with CMB and SNe~Ia datasets (Pantheon$+$, DES-Dovekie, and Union3).}\label{tab_2}
\end{table*}

\section{Results}\label{sec_4}
Fig.~\ref{fig_3} shows the posterior distributions of the parameters of the TPDE model at the 68\% ($1\sigma$) and 95\% ($2\sigma$) confidence regions, obtained by using the combinations of CMB + DESI DR2, CMB + DESI DR2 + Pantheon$+$, CMB + DESI DR2 + DES-Dovekie, and CMB + DESI DR2 + Union3. The off-diagonal panels show the two-dimensional (2D) marginalized joint posterior distributions for each pair of cosmological parameters, while the diagonal panels show the 1D marginalized posterior distributions for the corresponding cosmological parameters. The corresponding mean values and uncertainties at the 68\% ($1\sigma$) confidence level are presented in Table~\ref{tab_2}.

First, we discuss the status of the $H_0$ tension in the case of the TPDE model after DESI DR2. To provide a complete solution to the $H_0$ tension, as shown by \citep{knox2020hubble}, one must first note that there is a negative correlation between the Hubble constant, $H_0$, and the sound horizon at the drag epoch, $r_d$. This correlation implies that if the value of $H_0$ increases, the value of $r_d$ decreases, and vice versa. Therefore, in order to obtain a higher value of $H_0$ close to the Riess measurement, $H_0=(73.04\pm1.04),\mathrm{km,s^{-1},Mpc^{-1}}$ \citep{riess2022comprehensive,Murakami:2023xuy,Breuval:2024lsv}, the sound horizon $r_d$ must be reduced by approximately $7\%$. Furthermore, it is well known that changing the dark energy equation of state from $w(z)=-1$ ($\Lambda$CDM) to $w(z)>-1$ (quintessence) leads to a lower inferred value of $H_0$, whereas $w(z)<-1$ (phantom-like) leads to a higher inferred value of $H_0$ \citep{Vagnozzi:2018jhn,Vagnozzi:2019ezj,Alestas:2020mvb,Banerjee:2020xcn,Lee:2022cyh,Colgain:2025nzf}.

The TPDE model belongs to the class of dynamical dark energy models. In this kind of model, the dark energy component becomes relevant only during the post-recombination history of the Universe. Consequently, the pre-recombination expansion history remains essentially unchanged, so one cannot decrease the value of $r_d$~\citep{Bernal:2016gxb,Aylor:2018drw,Knox:2019rjx,Efstathiou:2021ocp,Jiang:2024xnu}. Indeed, for all dataset combinations, the inferred values of $r_d$ remain nearly unchanged, showing deviations about $0.45\sigma$, $0.84\sigma$, $0.87\sigma$, and $0.78\sigma$ for CMB + DESI DR2, CMB + DESI DR2 + Pantheon$^+$, CMB + DESI DR2 + DES-Dovekie, and CMB + DESI DR2 + Union3, respectively, when compared with the Planck 2018 determination, $r_d=147.21\pm0.23~\mathrm{Mpc}$ \citep{aghanim2020planck}. While the TPDE model predicts a relatively higher value of $H_0$, it does not provide a complete resolution of the Hubble tension, as $r_d$ remains unchanged. The inferred values correspond to deviations of approximately $3.19\sigma$, $3.98\sigma$, $4.18\sigma$, and $3.77\sigma$ with respect to the Riess measurement, $H_0=(73.04\pm1.04),\mathrm{km,s^{-1},Mpc^{-1}}$ \citep{riess2022comprehensive,Murakami:2023xuy,Breuval:2024lsv}, for the CMB + DESI DR2, CMB + DESI DR2 + Pantheon$^+$, CMB + DESI DR2 + DES-Dovekie, and CMB + DESI DR2 + Union3 dataset combinations, respectively.

Overall, one can conclude that, although the class of dynamical dark energy models provides a good description of the combined BAO+CMB+SNe data, it does not resolve the $H_0$ tension. In other words, dynamical dark energy fails to simultaneously fit BAO+CMB+SNe together with the local $H_0$ measurement.\citep{Verde:2023lmm}. An alternative strategy for lowering the sound horizon, $r_d$, is to modify the expansion history of the Universe during the pre-recombination era, roughly spanning $z\sim10^4$ to $z\sim10^3$. Such modifications typically require an enhancement of the physical matter density, as is the case in Early Dark Energy (EDE) scenarios~\citep{poulin2019early}. While EDE models can provide a good fit to the combined BAO and CMB data and help ease the $H_0$ tension, they tend to fit the SNe~Ia data less well and often worsen the existing $S_8$ tension~\citep{chaussidon2025early}.

We also investigate the clustering parameter $S_8 \equiv \sigma_8\left(\Omega_m/0.3\right)^{0.5}$ in the TPDE model, as it provides a key diagnostic of the growth of large-scale structure and the so-called $S_8$ tension. Current weak-lensing measurements, such as DES~Y6 ($S_8 = 0.794^{+0.009}_{-0.012}$)~\citep{DES:2026fyc} and KiDS-Legacy (Shear) ($S_8 = 0.815^{+0.016}_{-0.021}$)~\citep{wright2025kids}, favor lower values than the combined Planck + ACT + SPT CMB determination, $S_8 = 0.836^{+0.012}_{-0.013}$~\citep{DES:2026fyc}. These measurements differ from the CMB result by approximately $2.4\sigma$ and $0.9\sigma$, respectively. A recent overview of $S_8$ measurements and their tensions with the combined Planck + ACT + SPT CMB constraint can be found in Table~1 of Ref.~\citep{pantos2026status}.

In our analysis, we compare the inferred $S_8$ values with the recent weak-lensing measurements from KiDS-Legacy (Shear) and DES~Y6 to assess the consistency of the TPDE model with current large-scale structure observations. Relative to the KiDS-Legacy (Shear) measurement, the CMB + DESI DR2, CMB + DESI DR2 + Pantheon$^+$, CMB + DESI DR2 + DES-Dovekie, and CMB + DESI DR2 + Union3 dataset combinations shows tensions of only $0.04\sigma$, $0.17\sigma$, $0.18\sigma$, and $0.14\sigma$, respectively. The corresponding tensions with the DES~Y6 measurement are $1.47\sigma$, $1.26\sigma$, $1.24\sigma$, and $1.30\sigma$, respectively. These results indicate that the TPDE model is in good agreement with the KiDS-Legacy (Shear) constraints and remains fully compatible with the DES~Y6 weak-lensing measurements, showing no significant $S_8$ tension.

The parameter $q_{de,0}$ characterizes the present-day deceleration associated with the dark-energy component and determines its current dynamical behavior. In particular, $q_{de,0}=-1$ corresponds exactly to the $\Lambda$CDM limit, while values below (above) $-1$ indicate phantom-like (quintessence-like) dark energy. For the CMB + DESI DR2 dataset, the best-fit value shows a mild preference for phantom-like behavior, with $q_{de,0}=-1.046\pm0.031$. However, after including the Pantheon$+$, DES-Dovekie, and Union3 supernova compilations, the constraints shift to $q_{de,0}=-1.003\pm0.022$, $-0.996\pm0.020$, and $-1.008^{+0.027}_{-0.024}$, respectively, all of which are fully consistent with the $\Lambda$CDM prediction within $1\sigma$. These results indicate that current observations provide no statistically significant evidence for a deviation from a cosmological constant, although the CMB + DESI DR2 combination shows a weak preference for phantom-like dark energy.

The constraints on $w_{de,0}$ are fully consistent with those obtained for $q_{de,0}$, as the present-day dark-energy equation-of-state parameter is directly related to the deceleration parameter through $w_{de,0}=(2q_{de,0}-1)/3$. The corresponding constraints, $w_{de,0}=-1.031\pm0.020$, $-1.002\pm0.015$, $-0.998\pm0.014$, and $-1.005^{+0.018}_{-0.016}$ for the CMB + DESI DR2, CMB + DESI DR2 + Pantheon$^+$, CMB + DESI DR2 + DES-Dovekie, and CMB + DESI DR2 + Union3 dataset combinations, respectively, reinforce the same physical picture. The CMB + DESI DR2 dataset shows a mild preference for phantom-like dark energy ($w_{de,0}<-1$), whereas the inclusion of each supernova compilation shifts the constraints toward $w_{de,0}\simeq-1$. Consequently, all dataset combinations remain consistent with the cosmological constant scenario within the $1\sigma$ confidence level, providing no evidence for dynamical dark energy in the TPDE model at the present epoch.

The maximum-likelihood comparison shows no statistically significant improvement of the TPDE model over $\Lambda$CDM. For the CMB + DESI DR2 combination, we obtain $\Delta\chi^{2}_{\rm MAP}=0.39$, showing a marginally better best fit for $\Lambda$CDM. After including Pantheon$+$, we find $\Delta\chi^{2}{\rm MAP}=0.57$, again showing a slight preference for $\Lambda$CDM. For CMB + DESI DR2 + DES-Dovekie, we find $\Delta\chi^{2}_{\rm MAP}=2.02$, while the CMB + DESI DR2 + Union3 combination gives $\Delta\chi^{2}_{\rm MAP}=0.96$, both mildly favoring $\Lambda$CDM at the level of the minimum $\chi^{2}$. Overall, the differences in the minimized $\chi^{2}$ values are small, indicating that TPDE and $\Lambda$CDM provide statistically comparable best fits to the data.

In contrast, the logarithmic Bayes factor favors the TPDE model over $\Lambda$CDM for all dataset combinations. Specifically, we obtain $\ln B_{\Lambda \rm CDM,TPDE}=-2.88$ for CMB + DESI DR2, indicating a moderate preference for TPDE. The inclusion of Pantheon$+$, DES-Dovekie, and Union3 yields $\ln B_{\Lambda \rm CDM,TPDE}=-1.36$, $-1.04$, and $-1.49$, showing a weak preference for the TPDE model according to the Jeffreys scale. Although the best-fit $\chi^{2}$ values are comparable between the two models, the Bayesian evidence consistently favors TPDE because it evaluates the average performance over the full prior-weighted parameter space rather than only the single maximum-likelihood solution.

In Fig.~\ref{fig_4}, we present the reconstructed evolution of $H(z)/(1+z)$ (first row), $q(z)$ (second row), $w(z)$ (third row), $f_{\mathrm{DE}}(z)$ (fourth row), and $c_s^2(z)$ (fifth row) as functions of redshift for the TPDE and $\Lambda$CDM models. In each panel, the solid curves represent the mean reconstructed evolution, while the darker and lighter shaded regions correspond to the $1\sigma$ and $2\sigma$ confidence intervals, respectively. The first row shows the reconstructed evolution of $H(z)/(1+z)$. For the CMB + DESI DR2 dataset, the TPDE model predicts a slightly lower expansion history than $\Lambda$CDM over the intermediate-redshift range, although the two models remain in good agreement within the quoted uncertainties. After including the Pantheon$+$, DES-Dovekie, and Union3 supernova samples, the reconstructed expansion histories become nearly indistinguishable, with only subtle residual differences visible in the zoomed-in panels.

The second row presents the evolution of the deceleration parameter, $q(z)$. Both TPDE and $\Lambda$CDM shows the expected transition from a decelerating ($q>0$) to an accelerating ($q<0$) Universe at similar redshifts. The reconstructed $q(z)$ curves are nearly identical for all dataset combinations, with only a marginal tendency for the TPDE model to predict a slightly less negative present-day deceleration parameter than $\Lambda$CDM. These differences are small and remain well within the corresponding $1\sigma$ confidence regions. Overall, both models are in good agreement with current cosmological observations, consistently describing the present-day accelerated expansion of the Universe.

The third row presents the reconstructed evolution of the dark energy equation of state, $w_{\rm de}(z)$, for the TPDE model, while the $\Lambda$CDM model corresponds to the constant value $w=-1$. The reconstruction obtained from the CMB + DESI DR2 dataset shows the largest departure from the cosmological constant, favoring a phantom-like equation of state ($w_{\rm de}<-1$) throughout the reconstructed redshift range, together with a noticeable late-time evolution. The inclusion of the Pantheon$+$ supernova sample substantially suppresses this evolution, yielding a reconstruction that is fully consistent with a cosmological constant, with $w_{\rm de}(z)\simeq-1$ across the entire redshift range. In contrast, the DES-Dovekie compilation shows a mild preference for a quintessence-like evolution ($w_{\rm de}>-1$), whereas the Union3 compilation shows a slight preference for the phantom regime ($w_{\rm de}<-1$). Nevertheless, both reconstructions remain close to $w=-1$ and approach the cosmological constant behaviour over the observed redshift range, indicating that any departure from $\Lambda$CDM is modest and lies within the reconstructed confidence regions.

The fourth row presents the reconstructed evolution of the normalized dark energy density, $f_{\rm DE}(z)$, for the TPDE model. By construction, all dataset combinations satisfy $f_{\rm DE}(0)=1$ at the present epoch, matching the $\Lambda$CDM prediction. The CMB + DESI DR2 dataset shows the strongest departure from the cosmological constant, with $f_{\rm DE}(z)$ decreasing monotonically below unity as the redshift increases, indicating that the contribution of dark energy gradually decreases at earlier cosmic times. In contrast, the inclusion of the Pantheon$+$ supernova sample suppresses this evolution, yielding a reconstruction that remains nearly indistinguishable from the constant $\Lambda$CDM prediction throughout the observed redshift range. The DES-Dovekie compilation shows a mild enhancement of the normalized dark energy density above the $\Lambda$CDM prediction, whereas the Union3 compilation shows a slight suppression below unity. Nevertheless, both reconstructions remain close to $f_{\rm DE}(z)=1$ within the reconstructed confidence regions, indicating that any departure from a cosmological constant is modest.

The fifth row presents the reconstructed evolution of the effective adiabatic sound speed squared, $c_a^2(z)$, of the total cosmic fluid for the TPDE model. The CMB + DESI DR2 dataset favors positive values of $c_a^2(z)$ throughout the reconstructed redshift range, gradually decreasing towards zero as the redshift increases. The inclusion of the Pantheon$+$ supernova sample strongly suppresses this evolution, yielding a reconstruction that remains nearly indistinguishable from $c_a^2=0$ over the entire redshift range. The DES-Dovekie compilation shows a slight preference for negative values of $c_a^2$ at low redshift, while the Union3 compilation favors small positive values. In both cases, however, the reconstructed evolution rapidly approaches $c_a^2=0$ with increasing redshift and remains fully consistent with vanishing effective adiabatic sound speed within the reconstructed confidence regions. Overall, the supernova datasets substantially tighten the allowed evolution of the effective adiabatic sound speed.

\section{Discussion and Conclusions}\label{sec_5}
In this work, we have investigated a thermodynamically motivated reconstruction of dark energy based on the fundamental principles of cosmological thermodynamics and the kinematic evolution of the Universe. Instead of introducing an arbitrary equation-of-state parameterization, we imposed the second law of thermodynamics together with physically motivated asymptotic conditions on the cosmic deceleration parameter. This procedure naturally led to a Padé reconstruction of the dark energy sector, resulting in a simple one-parameter extension of the standard $\Lambda$CDM cosmology. Consequently, the proposed TPDE model provides a physically motivated alternative to the cosmological constant, where the late-time accelerated expansion emerges from thermodynamic considerations rather than from an ad hoc phenomenological description.

We constrained the model using the combination of Planck 2018 legacy CMB temperature and polarization observations, Planck 2018 CMB lensing, DESI DR2 BAO measurements, and three different SNe~Ia compilations (Pantheon$+$, DES-Dovekie, and Union3). The reconstructed cosmological evolution demonstrates that the TPDE model successfully reproduces all the main features of the observed expansion history, including the transition from a decelerating matter-dominated Universe to the present epoch of cosmic acceleration. The reconstructed Hubble expansion rate, deceleration parameter, dark energy equation of state, normalized dark energy density, and effective adiabatic sound speed remain in good agreement with the predictions of the standard $\Lambda$CDM model throughout the observationally accessible redshift range.

The cosmological constraints show that the present-day thermodynamic parameter is tightly bounded around $q_{\rm de,0}\simeq-1$. Although the CMB + DESI DR2 dataset alone shows a mild preference for phantom-like dark energy, the inclusion of each of the supernova compilations drives the constraints towards the $\Lambda$CDM limit, indicating that current observations provide no statistically significant evidence for departures from a cosmological constant. The corresponding reconstruction of the dark energy equation of state likewise remains fully consistent with $w=-1$ within the $1\sigma$ confidence level, implying that any possible dynamical evolution of dark energy is strongly constrained by present cosmological observations.

We also investigated the cosmological implications of the TPDE model for the current observational tensions. Since the proposed scenario modifies only the late-time expansion history, the sound horizon at the baryon drag epoch remains essentially unchanged. Consequently, although the model predicts slightly higher values of the Hubble constant than the standard cosmological model, it does not provide a complete solution to the $H_0$ tension. This result is consistent with the general expectation that late-time dark energy models alone cannot simultaneously reconcile CMB, BAO, SNe~Ia observations, and local determinations of the Hubble constant. In contrast, the TPDE model shows good agreement with current weak-lensing measurements of the clustering parameter $S_8$. Relative to the KiDS-Legacy (Shear) measurements, all dataset combinations shows tensions below $0.2\sigma$, while the corresponding tensions with the DES~Y6 weak-lensing constraints remain below $1.5\sigma$. Therefore, the TPDE scenario remains fully consistent with current large-scale structure observations and does not introduce any significant $S_8$ tension.

From a statistical perspective, the TPDE model provides fits to the observational data that are comparable to those of $\Lambda$CDM at the level of the maximum likelihood. However, the Bayesian evidence consistently yields a weak-to-moderate preference for the TPDE model over $\Lambda$CDM across all dataset combinations. These results indicate that a thermodynamically reconstructed dark energy scenario provides a competitive and physically motivated alternative to the standard cosmological model, demonstrating that fundamental thermodynamic principles can successfully describe the late-time accelerated expansion of the Universe while remaining observationally favored over the conventional $\Lambda$CDM scenario according to Bayesian model comparison.

Although present observations do not require significant deviations from a cosmological constant, the TPDE framework offers an attractive and theoretically well-motivated alternative to the standard $\Lambda$CDM paradigm. Unlike many commonly adopted phenomenological dark energy parameterizations, the proposed reconstruction is directly rooted in the thermodynamic properties of the Universe and therefore possesses a clear physical interpretation. As future cosmological surveys such as Euclid, the Vera C. Rubin Observatory Legacy Survey of Space and Time (LSST), the Nancy Grace Roman Space Telescope, DESI DR3, and CMB-S4 substantially improve the precision of cosmological measurements, they will enable much more stringent tests of thermodynamically reconstructed dark energy scenarios. Such observations will determine whether dark energy is indeed a cosmological constant or whether subtle departures from the standard $\Lambda$CDM picture emerge from the fundamental thermodynamic evolution of the Universe.

\section*{Acknowledgements}
SC acknowledges the Istituto Nazionale di Fisica Nucleare (INFN) Sez. di Napoli,  Iniziative Specifiche QGSKY and MoonLight-2  and the Istituto Nazionale di Alta Matematica (INdAM), gruppo GNFM, for the support. This paper is based upon work from COST Action CA21136 -- Addressing observational tensions in cosmology with systematics and fundamental physics (CosmoVerse), supported by COST (European Cooperation in Science and Technology). This research was funded by the National Natural Science Foundation of China (NSFC) under Grant No. U2541210 and 12505238.

\section*{Data and Code Availability Statements}
All data used in this study are publicly available from the corresponding survey collaborations and repositories cited in the main text. The computational codes and software employed in the analysis are publicly available, and their corresponding GitHub repositories or official source links are provided in the relevant citations throughout the manuscript. 

\bibliographystyle{aa}
\bibliography{ref}

\end{document}